\documentclass[twocolumn,showpacs,preprintnumbers,amsmath,amssymb]{revtex4}

\usepackage{graphicx}
\usepackage{dcolumn}
\usepackage{bm}

\begin{document}


\title{Spin transport through a single self-assembled InAs quantum dot \\with ferromagnetic leads}

\author{K. Hamaya\footnote{hamaya@iis.u-tokyo.ac.jp}, S. Masubuchi, M. Kawamura, and T. Machida\footnote{tmachida@iis.u-tokyo.ac.jp}}
\affiliation{%
Department of Fundamental Engineering, Institute of Industrial Science, University of Tokyo, 4-6-1 Komaba, Meguro-ku, Tokyo 153-8505, Japan
}%

\author{M. Jung, K. Shibata, and K. Hirakawa}
\affiliation{%
Department of Informatics and Electronics, Institute of Industrial Science, University of Tokyo, 4-6-1 Komaba, Meguro-ku, Tokyo 153-8505, Japan 
}%

\author{T. Taniyama}
\affiliation{%
Materials and Structures Laboratory, Tokyo Institute of Technology, 4259 Nagatsuta, Midori-ku, Yokohama 226-8503, Japan 
}%

\author{S. Ishida and Y. Arakawa}
\affiliation{%
Nanoelectronics Collaborative Research Center (NCRC), IIS and RCAST, University of Tokyo, 4-6-1 Komaba, Meguro-ku, Tokyo 153-8505, Japan }

%


\date{\today}

\begin{abstract}
We have fabricated a lateral double barrier magnetic tunnel junction (MTJ) which consists of a single self-assembled InAs quantum dot (QD) with ferromagnetic Co leads. The MTJ shows clear hysteretic tunnel magnetoresistance (TMR) effect, which is evidence for spin transport through a single semiconductor QD. The TMR ratio and the curve shapes are varied by changing the gate voltage. 
\end{abstract}

\maketitle

 
The research field of semiconductor-based spin electronics (spintronics) has opened up a new technology for spin manipulation by means other than magnetic field.\cite{Kato,Ohno} For developing semiconductor nanospintronic applications and discovering novel physical phenomena, one is extremely interested in technological possibilities for spin injection into a single semiconductor quantum dot (QD) which behaves as an artificial atom.\cite{Tarucha} To date, many theoretical studies of spin transport through a single nonmagnetic island with ferromagnetic leads have been reported,\cite{Brataas,Imamura,Jan,Rud,Weymann,Wetzels} and spin accumulation in the island was predicted in their reports. Very recently, for metallic systems, spin injection into a single nonmagnetic nanoparticle was achieved,\cite{Fert} which indicates the occurrence of spin accumulation. For an individual carbon nanotube (CNT) with ferromagnetic leads, the spin transport\cite{Tsukagoshi,Zhao} and its gate-control\cite{Sahoo,Jensen,Nagabhirava,Man} have also been demonstrated, showing possible spintronic applications using CNTs. However, no experimental work on spin-dependent transport through a single semiconductor QD has been reported yet.  

Recently, Jung {\it et al.}\cite{Jung} succeeded in transport measurements for a single self-assembled InAs QD in contact with nonmagnetic leads and clearly observed shell structures due to an artificial atomic nature. Replacing the nonmagnetic leads with ferromagnetic ones, we can inject spin-polarized electrons from the ferromagnetic leads into a single InAs QD. A number of works proposed that if the spin relaxation time in the QD is sufficiently long, spin accumulation can occur, which is detected through tunneling magnetoresistance (TMR) effects.\cite{Brataas,Imamura,Jan,Rud,Weymann,Wetzels,Mattana,Yakushiji} In semiconductor heterostructures, the electrical detection of the spin accumulation in the double barrier magnetic tunnel junction (MTJ), (Ga,Mn)As/AlAs/GaAs/AlAs/(Ga,Mn)As, has been demonstrated by measuring the TMR effect.\cite{Mattana}
 \begin{figure}[t]
\includegraphics[width=6.5cm]{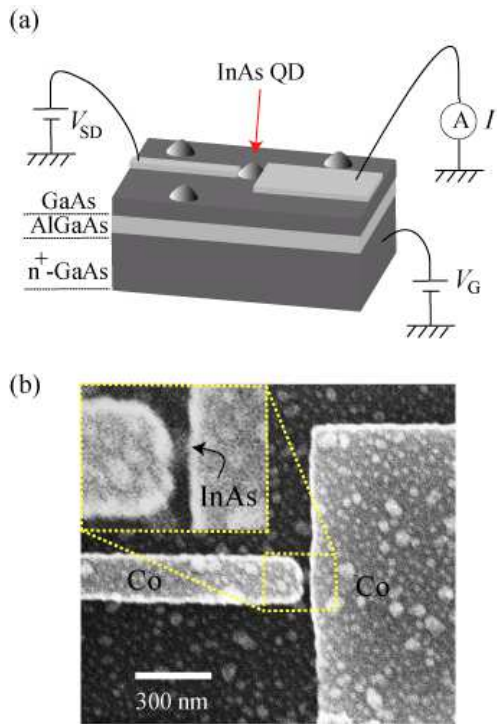}
\caption{(a) Schematic diagram of a Co/InAs/Co double barrier magnetic tunnel junction. (b) Scanning electron micrograph of the Co/InAs/Co junction structure used in the present study. The inset shows an enlarged micrograph in the enclosed area.}
\end{figure}

In this letter, we report on the observation of spin transport through a single semiconductor QD using a lateral double barrier MTJ composed of ferromagnetic Co leads and a single self-assembled InAs QD. The MTJ shows clear Coulomb blockade effects and the TMR effects. The TMR features are varied evidently by changing the gate voltage, meaning that the spin transport can be tuned by applying the electric field. 
  
Self-assembled InAs QDs were grown on a substrate made of, from the top, 200-nm-thick GaAs buffer layer, 120-nm-thick AlGaAs insulating layer, and $n$$^{+}$-GaAs(001). The $n$$^{+}$-GaAs(001) was used as a backgate electrode.  Using electron-beam lithography and lift-off method, we fabricated the wire-shape Co leads with a $\sim$ 30-nm gap. Our device structure is a lateral MTJ, schematically illustrated in Fig.1 (a). A single InAs QD is in contact with two Co wires that have a thickness of 40 nm, deposited by means of an ultrahigh vacuum evaporator with a base pressure of 1 $\times$ 10$^{-9}$ Torr. In order to induce asymmetric shape anisotropy, one of the Co leads was formed to be 1 $\mu$m-wide and 20 $\mu$m-long, while the other was formed to be 200 nm-wide and 20 $\mu$m-long, giving rise to different switching fields for each lead. Since there exists a natural oxide layer formed on the InAs surface,\cite{ref} the Co/InAs interface layers act as tunnel barriers. Thus, the device used in this study is a lateral double barrier MTJ, Co/InAs/Co. A scanning electron micrograph of the Co/InAs/Co MTJ is shown in Fig. 1(b). A single InAs QD with a size of $\sim$ 80 nm is located between the two Co leads. Transport measurements were performed by the dc method in a $^{3}$He$-$$^{4}$He dilution refrigerator at $\sim$50 mK.
\begin{figure}[t]
\includegraphics[width=8cm]{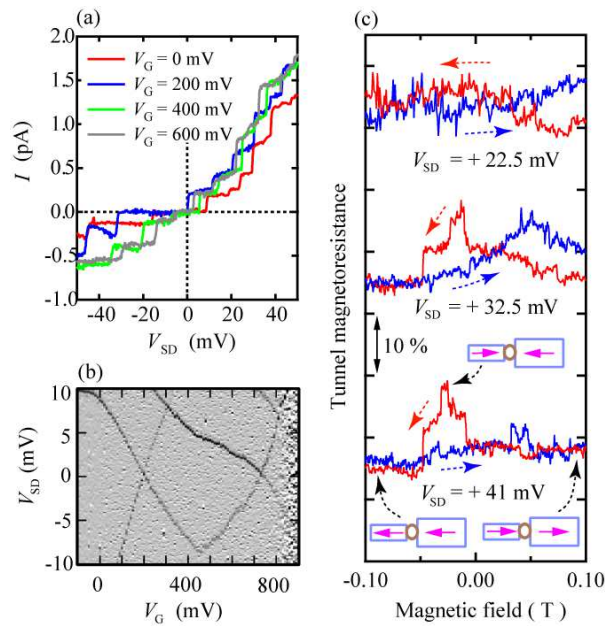}
\caption{(a) Current-voltage ($I-V_\mathrm{SD}$) characteristics of a Co/InAs/Co junction structure, measured at 50 mK, for various backgate voltages, $V$$_\mathrm{G}$. (b) Differential conductance, $dI$/$d$$V$$_\mathrm{SD}$, as functions of $V$$_\mathrm{SD}$ and $V$$_\mathrm{G}$. (c) Magnetic field dependence of tunnel resistance for various $V$$_\mathrm{SD}$. The inset shows schematic illustrations of the magnetic configuration of the leads.}
\end{figure} 

The current vs source-drain voltage ($I -$$V$$_\mathrm{SD}$) characteristics of the Co/InAs/Co MTJ at 50 mK are shown in Fig. 2(a) for various backgate voltages ($V$$_\mathrm{G}$). These curves are measured under an external magnetic field of $\sim$ 0.2 T parallel to the long axis of the Co leads, where the magnetizations of both leads become nearly parallel. The $I -$$V$$_\mathrm{SD}$ characteristics clearly show the zero conductance in the vicinity of $V$$_\mathrm{SD} =$ 0 V, indicating Coulomb blockade. Also, step-like $I -$$V$$_\mathrm{SD}$ structures, so-called Coulomb staircase, can be seen. It should be noted that the resistance jumps of $\sim$ G$\Omega$ are much larger than the quantum resistance of $h$/$e$$^{2}$ $\sim$ 25.8 k$\Omega$, and the junction resistance is very large compared to the previous studies by Jung {\it et al.}\cite{Jung} who used nonmagnetic Au leads. Thus, the couplings of the dot to the Co leads are very weak and sequential tunneling processes are likely to be dominant for our sample. Furthermore, the $I -$$V$$_\mathrm{SD}$ curves show asymmetric features about the polarity of $V$$_\mathrm{SD}$. This means that in our device the two tunnel barriers is asymmetric. In Fig. 2(b) we also measure the differential conductance, $dI$/$d$$V$$_\mathrm{SD}$, as functions of $V$$_\mathrm{SD}$ and $V$$_\mathrm{G}$. Unfortunately, since the accessible range of the $V$$_\mathrm{G}$ is very narrow, we observe only one diamond-like shape of the $dI$/$d$$V$$_\mathrm{SD}$ in Fig. 2(b). However, these operations verify that our device is working as a single electron transistor, as well as nonmagnetic Au leads.\cite{Jung} We also emphasize that this is the demonstration of Coulomb blockade effects for a single semiconductor QD with ferromagnetic leads.
\begin{figure}
\includegraphics[width=8.5cm]{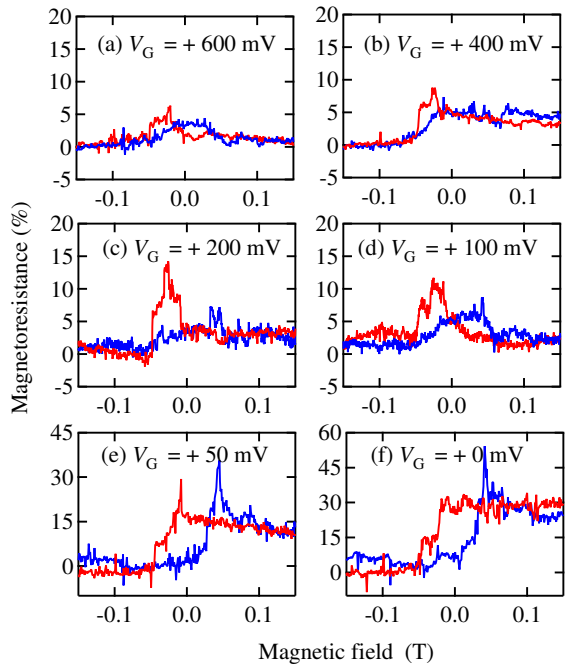}
\caption{(a) Tunneling magnetoresistance curves of a Co/InAs/Co junction measured at $V$$_\mathrm{SD} =$ 41 mV at 50 mK for (a) $V$$_\mathrm{G} =$ 600 mV, (b) 400 mV, (c) 200 mV, (d) 100 mV, (e) 50 mV, and (f) 0 mV.}
\end{figure}     

Figure 2(c) shows the magnetic field dependence of the tunnel resistance for $V$$_\mathrm{SD} =$ 22.5, 32.5, and 41 mV at $V$$_\mathrm{G}$ = 200 mV, measured at 50 mK. The magnetic field ($H$) is applied parallel to the long axis of the wires. The data traces are recorded for two sweep directions of the magnetic field: the blue curves are up-sweep from $-$ 0.2 to + 0.2 T, while the red curves are down-sweep from + 0.2 to $-$ 0.2 T. Here, the TMR ratio (\%) is defined as $\{$($I_\mathrm{H} -$ $I_\mathrm{-0.1 T}$)/$I_\mathrm{- 0.1T}$$\}$ $\times$ 100, where $I_\mathrm{H}$ and $I_\mathrm{- 0.1T}$ are the tunnel currents for magnetic field of $H$ and for $H =$ $-$ 0.1 T, respectively. First, we focus on the data for $V$$_\mathrm{SD} =$ 41 mV. When the magnetic field is swept from + 0.2 T toward negative fields (red curve), the magnetizations of the two Co leads are changed from a parallel configuration into an approximately anti-parallel configuration, as shown in the inset of Fig. 2(c), and the tunnel resistance becomes large at $H\approx$ $-$ 0.02 T. Similar features are seen in up-sweep procedure (blue curve). This behavior is a positive TMR effect. The TMR curve having typical hysteretic features is also observed for $V$$_\mathrm{SD} =$ 32.5 mV. These TMR curves clearly suggest that electron spins are conserved in the tunneling transport processes. These are the observation of spin-dependent transport through a single semiconductor QD with ferromagnetic leads. On the other hand, the TMR signal can not be clearly seen at $V$$_\mathrm{SD} =$ 22.5 mV. For our device the spin transport is detected only at higher $V$$_\mathrm{SD}$ regime, and it is difficult to discuss the detailed correlation between $I -$$V$$_\mathrm{SD}$ characteristics and TMR features.
 
In Figs. 3(a)-(f), we show the TMR curves for various $V$$_\mathrm{G}$ measured at $V$$_\mathrm{SD} =$ 41 mV. We note that the magnitude of the TMR ratio and the curve shape can be manipulated by changing $V$$_\mathrm{G}$. From $V$$_\mathrm{G} =$ 600 to 100 mV [Figs. 3(a)-(d)], the change in the TMR ratio is small but slight variation in the shape of the TMR curve is seen with decreasing $V$$_\mathrm{G}$. These features are reproduced qualitatively. Surprisingly, the shape of the TMR curve and the TMR ratio change markedly at $V$$_\mathrm{G} =$ 50 and 0 mV [Figs. 3(e) and 3(f)]. 

One of the possible mechanisms of the spin-dependent transport presented here is spin accumulation in a single InAs QD by injecting spin-polarized electrons from one ferromagnetic Co lead.\cite{Brataas,Imamura,Jan,Rud,Weymann,Wetzels,Mattana,Yakushiji} In theoretical studies based on the spin accumulation,\cite{Brataas,Imamura,Jan,Rud,Weymann,Wetzels} the bias dependence and the gate-voltage dependence of the TMR ratio have been predicted. By means of optical measurements, the spin relaxation time in self-assembled InAs QDs was deduced to be $\sim$1 ns at low temperature,\cite{Takeuchi} which is long enough to induce a nonequilibrium spin accumulation. However, the unexpected large TMR ratio, shown in Figs. 3(e) and 3(f), can not currently be explained by previously reported theoretical predictions.\cite{Brataas,Imamura,Jan,Rud,Weymann} Also, the origin of the shape change in the TMR curve is unclear yet. In this regard, we speculate that the influence of complicated domain structures in the wider lead  and/or the magneto-Coulomb effect\cite{Molen} on the tunnel conductance should also be taken into account.

Recently, for PdNi/CNT/PdNi systrems,\cite{Sahoo} the gate-dependent TMR was reported, in which the dependence may originate from the discrete energy level depending on the gate voltage and the spin orientation of the ferromagnetic leads. In our case, the influence of the discrete levels in the QD, which may have the spin splitting due to spin accumulation,\cite{Brataas,Imamura,Jan,Rud,Weymann,Wetzels} should also be considered. 

In summary, we have fabricated a double barrier MTJ which consists of a single self-assembled InAs QD in contact with ferromagnetic Co leads. Clear Coulomb blockade effects and TMR effects are demonstrated and they can be varied by changing the gate bias voltage. 

\vspace{2mm}
K. H. and T. M. thank Prof. S. Tarucha and Dr. J. Martinek for helpful discussions. This work is supported by the Special Coordination Funds for Promoting Science and Technology, and Collaborative Research Project of Materials and Structures Laboratory, Tokyo Institute of Technology. K. H. acknowledges JSPS Research Fellowships for Young Scientists. 


\begin{thebibliography}{11}
\bibitem{ref}
In this study, the tunnel resistance is quite large ($\sim$ G$\Omega$) for all the samples studied. Thus, the magnitude of the conductivity seems to be dominated by tunnel barriers formed at the ferromagnet/semiconductor interface.
\end{thebibliography}
\end{document}